\documentclass[cameraready]{Interspeech}
\title{Uni-ASR: Unified LLM-Based Architecture for Non-Streaming and Streaming Automatic Speech Recognition}
\author[affiliation={1}, orcid=0000-0002-4638-4009]{Yinfeng}{Xia}
\author[affiliation={2}, orcid=0000-0002-6819-4185]{Jian}{Tang}
\author[affiliation={2}, orcid=0000-0003-3635-5332]{Junfeng}{Hou}
\author[affiliation={1}, orcid=0009-0008-5980-2127]{Gaopeng}{Xu}
\author[affiliation={1}, correspondingauthor]{Haitao}{Yao}
\address{
    $^1$ Qwen Applications Business Group, Alibaba, China \\
    $^2$ Tongyi AI Lab, Alibaba, China
}

\email {\{xiayinfeng.xyf,yongran.tj,qianzhuo.hjf,longque.xgp,timmy.yht\}@alibaba-inc.com}
\keywords{automatic speech recognition, streaming ASR}

\usepackage{fontawesome5}
\usepackage{xcolor}
\usepackage{comment}
\usepackage{subfigure}
\usepackage{graphicx}
\usepackage{subcaption}
\usepackage{boldline}
\usepackage{wasysym}
\usepackage{multirow}
\usepackage{nicematrix}
\usepackage{makecell}
\usepackage{booktabs}
\setcellgapes{0.5pt} % 设置单元格上下留白
\makegapedcells   % 启用留白

\begin{document}

\maketitle

\begin{abstract}
Although the deep integration of the Automatic Speech Recognition (ASR) system with Large Language Models (LLMs) has significantly improved accuracy, the deployment of such systems in low-latency streaming scenarios remains challenging. In this paper, we propose Uni-ASR, a unified framework based on LLMs that integrates both non-streaming and streaming speech recognition capabilities. We propose a joint training paradigm that enables the system to seamlessly transition between two recognition modes without any architectural modifications. Furthermore, we introduce a context-aware training paradigm and a co-designed fallback decoding strategy, which can enhance streaming recognition accuracy without introducing additional latency. The experimental results demonstrate that Uni-ASR not only achieves competitive performance within non-streaming mode, but also demonstrates strong effectiveness in streaming scenarios under diverse latency constraints.

% We propose a hybrid training paradigm that enables the trained speech recognition system to seamlessly transition between non-streaming and streaming decoding modes—without requiring any architectural modifications.

% We propose a hybrid training paradigm that enables both non-streaming and streaming speech recognition within a unified system. 

% We propose a novel training paradigm that aligns and interleaves speech representations with text tokens at the chunk-level to construct the input sequence during the training phase, mimicking how LLMs process speech streams during decoding.

%Although the deep integration of automatic speech recognition (ASR) with large language models (LLMs) has significantly improved accuracy, fundamental system limitations hinder their deployment in low-latency streaming scenarios.
\end{abstract}

\section{Introduction}

In recent years, breakthroughs in Automatic Speech Recognition (ASR) have been primarily driven by the synergistic scaling of training data and model parameters. End-to-end Attention-based Encoder-Decoder (AED) frameworks--Whisper~\cite{radford2023robust}, have validated the efficacy of scaling principles in advancing multilingual speech recognition, leveraging architectures with 1.5B parameters trained on 680k hours of multilingual data. Building upon this foundation, the ASR paradigm is rapidly evolving toward deep integration with Large Language Models (LLMs)~\cite{mu2025efficient}. Systems such as Seed-ASR~\cite{bai2024seed}, FireRedASR~\cite{xu2025fireredasr}, Fun-ASR~\cite{an2025fun}, Qwen3-ASR~\cite{shi2026qwen3} and Index-ASR~\cite{song2025index} have demonstrated marked superiority over conventional AED architectures, achieving substantial gains in semantic disambiguation, contextual coherence, and transcription fluency. This evolution underscores the emergence of LLM-based approaches as the dominant paradigm for state-of-the-art (SOTA) ASR. Nevertheless, deploying such systems in latency-sensitive scenarios, such as real-time captioning, remains a formidable challenge. To mitigate this constraint, current research efforts coalesce around two primary implementation paradigms.

The first paradigm construct a dedicated streaming decoding pipeline for conventional non-streaming ASR models, enabling incremental inference without architectural modifications or model retraining. Representative incremental inference strategies—such as hold-\textit{n}, wait-\textit{k} and local agreement~\cite{liu2020low}—selectively emit partial hypotheses rather than generating full chunk-level predictions. This design mitigates acoustic ambiguities near chunk boundaries and suppresses error propagation caused by low-confidence prefix tokens, thereby preserving overall transcription reliability. However, a training-inference mismatch introduces two critical inefficiencies during incremental decoding: (1) substantial computational redundancy incurred by repeated re-initialization of the decoding process, and (2) unpredictable latency spikes triggered by the hypothesis selection process. Collectively, these limitations compromise both decoding efficiency and real-time performance.

The second paradigm focuses on building dedicated ASR models with native streaming processing capabilities. Representative works include: Transducer-Llama~\cite{deng2025transducer} integrates a LLM into the Factorized Transducer framework as a non-blank predictor for contextual refinement; SpeechLLM-XL~\cite{jia2025efficient} and Speech ReaLLM~\cite{seide2024speech} divide speech-text sequences into segments for streaming training; MoCha-ASR~\cite{wan2026streaming} adopts monotonic chunk-wise attention with a read-write policy network and MinLT loss to reduce streaming latency. Nevertheless, these methods exhibit inherent limitations: Transducer-Llama treat LLM merely as an auxiliary component, lacking deep speech-language modality integration and thereby constraining end-to-end collaborative optimization; the experimental validation of SpeechLLM-XL and Speech ReaLLM is limited to small language models, and their scalability and effectiveness on LLMs have not been fully verified; MoCha-ASR introduces substantial complexity to the streaming inference pipeline despite its end-to-end optimization objectives.

In this paper, we present \textbf{Uni-ASR}, a unified LLM-based ASR framework that natively supports both non-streaming and streaming inference within a single architecture. Our joint training framework integrates the established non-streaming paradigm with the proposed streaming paradigm. Specifically, the streaming branch leverages external forced alignment to construct interleaved speech-text sequences, accompanied by loss masks tailored for incremental decoding. This formulation maintains the high recognition accuracy of non-streaming systems while accommodating the low-latency inference demands inherent to streaming applications. Furthermore, we introduce a context-aware streaming training paradigm coupled with a fallback decoding strategy, where input sequences and labels are deliberately biased to reflect real decoding constraints. This ensures training-inference consistency, significantly boosting streaming performance without adding inference delay. Extensive experiments on open-source ASR benchmarks demonstrate that Uni-ASR not only achieves non-streaming accuracy competitive with SOTA systems, but also delivers robust streaming performance across diverse latency budgets. Ablation studies confirmed the effectiveness of our proposed training paradigm and decoding strategy in streaming mode.

\clearpage

\section{Methods}
\subsection{Model Architecture}

\begin{figure} %[t]
  \centering
  \includegraphics[width=3in]{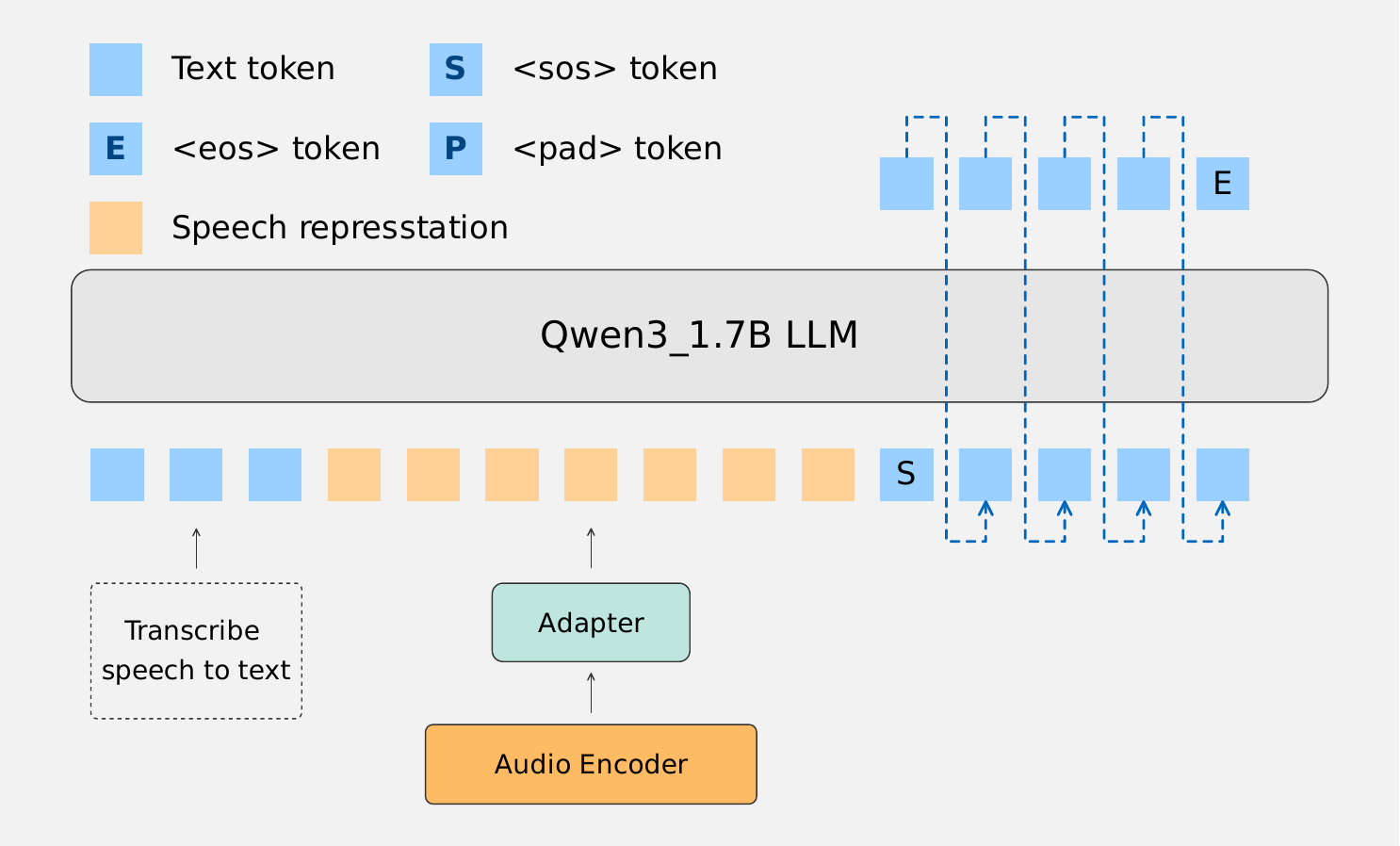}
  \caption{Overview of the Uni-ASR model architecture.}
  \label{fig1_architecture}
\end{figure}

Uni-ASR follows the architectural paradigm of established LLM-based ASR frameworks~\cite{bai2024seed,xu2025fireredasr,an2025fun,shi2026qwen3,song2025index}, comprising three core components: an audio encoder, an audio adapter, and an LLM-based decoder, as illustrated in Figure~\ref{fig1_architecture}.

% Uni-ASR follows the architectural paradigm of established LLM-based ASR frameworks.

% The audio encoder employs the Conformer~\cite{gulati2020conformer} architecture to extract high-level speech representations from the input speech. The audio adapter comprises two linear layers and a ReLU activation layer, projecting speech representation into the embedding dimension required by the LLM. The decoder adapts the pre-trained Qwen3-1.7B~\cite{yang2025qwen3} language model to autoregressively generate transcriptions, conditioned on encoder-derived speech representations and user-provided textual prompts.

The audio encoder employs the Conformer~\cite{gulati2020conformer} architecture to extract high-level speech representations from the input speech. The audio adapter—comprising two linear layers with an intermediate ReLU activation—projects these representations into an embedding space compatible with the LLM. The decoder leverages the pre-trained Qwen3-1.7B language model~\cite{yang2025qwen3} to autoregressively generate transcriptions, conditioned on both encoder-derived speech representations and user-provided textual prompts.

\subsection{Joint Training Paradigm and Dedicated Decoding Strategy}
\label{training paradigms}

We propose an integrated training-decoding framework for the Uni-ASR system, featuring a unified joint training paradigm paired with dedicated decoding strategies. This framework seamlessly supports both non-streaming and streaming inference within a single end-to-end architecture, eliminating the need for separate model variants.

% We introduce a unified joint training paradigm and a streaming decoding protocol for the Uni-ASR system, enabling seamless integration of non-streaming and streaming inference within a single, cohesive architecture.

% We present a cohesive training-decoding framework for the Uni-ASR system, comprising a unified joint training paradigm and its corresponding decoding strategy. This design seamlessly unifies non-streaming and streaming inference capabilities within a single, end-to-end architecture.

% We propose a joint training paradigm for the Uni-ASR system, leveraging deliberately designed input sequence configurations and tailored loss constraints to enable seamless support for both non-streaming and streaming inference.

\subsubsection{Non-streaming Paradigm}
In non-streaming ASR systems, the model processes the complete utterance as input to generate the transcription. Within the non-streaming~(\textbf{NS}) training paradigm employing a teacher-forced autoregressive framework, the input sequence is constructed by concatenating the speech representation, a start-of-sequence~(\verb+<sos>+) token , and the ground-truth text tokens in chronological order. The corresponding target sequence is formed by shifting the ground-truth tokens forward by one position and appending an end-of-sequence~(\verb+<eos>+) token, which ensures proper sequence termination during inference to avoid unbounded decoding. As illustrated in Figure~\ref{Fig2_1a}, during the minimization of cross-entropy loss, the loss is computed only over the target text tokens.

% For non-streaming inference, the entire speech representation sequence is fed into the LLM during the prefill stage. The decoding stage subsequently begins, where the text tokens are generated sequentially until the \verb+<eos>+ token is encountered, signaling the completion of the output sequence.

For non-streaming inference, the entire speech representation sequence is input into the LLM during the prefilling stage. Then, in the decoding stage, text tokens are generated sequentially until the token \verb+<eos>+ is encountered, indicating the completion of the output sequence.

% prediction targets—specifically, the tokens corresponding to the ground-truth transcription (excluding the speech representation prefix).

% For online recognition systems, we process the input as fixed-length speech blocks and use an external forced alignment tool to segment the text transcription. We treat the online training paradigm as a collection of consecutive offline training paradigms, meaning that the recognition process for any chunk can be viewed as a separate offline recognition process with contextual memory. In this manner, we construct speech-text interleaved input sequences and corresponding target labels for model training. Furthermore, referencing the training format of cosyvoice2, we mix speech and text tags in a predefined ratio of N:M, i.e., every N speech tags are followed by M text tags. To avoid the influence of speech rate and language, we use \verb+<pad>+ to pad excess text tags. 

\subsubsection{Standard Streaming Paradigm}
In streaming ASR systems, the input speech stream is divided into fixed-length chunks, and an external forced alignment tool is employed to align and segment the corresponding transcribed text. We formulate the streaming training process as multiple rounds of consecutive non-streaming training instances, wherein each speech chunk is processed as an independent recognition task enriched with contextual memory propagated from preceding segments. This standard streaming~(\textbf{SS}) paradigm facilitates end-to-end model training by constructing interleaved speech-text input sequences and pairing them with target transcriptions. 

\begin{figure} %[!hptb]
\centering
\subfigure[Non-streaming training paradigm]{\label{Fig2_1a}
\begin{minipage}[hptb]{\linewidth}
\centering
\includegraphics[width=3.0in]{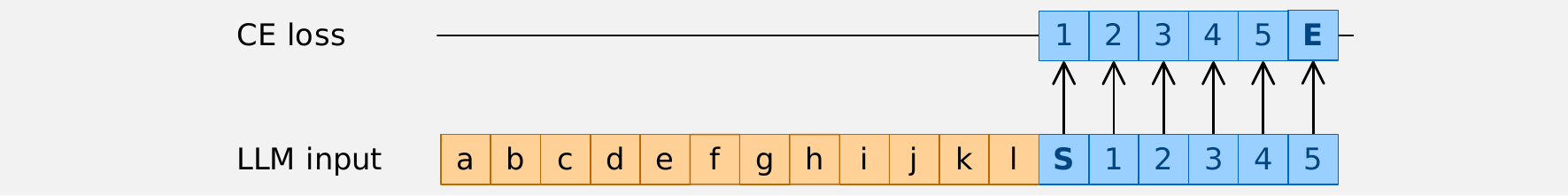}
\end{minipage}%
}%
\quad
\subfigure[Standard streaming training paradigm]{\label{Fig2_1b}
\begin{minipage}[hptb]{\linewidth}
\centering
\includegraphics[width=3.0in]{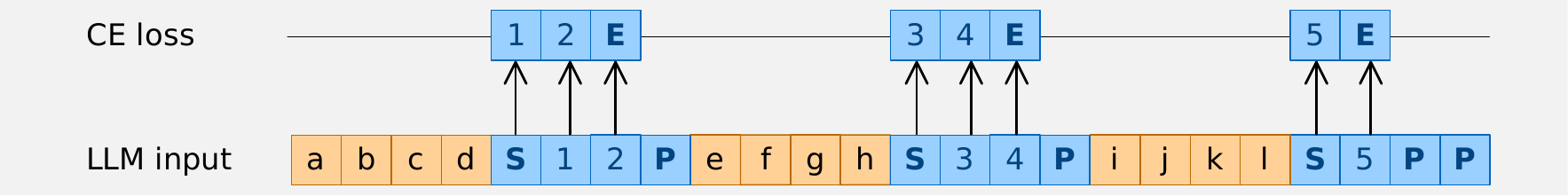}
\end{minipage}%
}%
\quad
\subfigure[Context-aware streaming training paradigm]{\label{Fig2_1c}
\begin{minipage}[hptb]{\linewidth}
\centering
\includegraphics[width=3.0in]{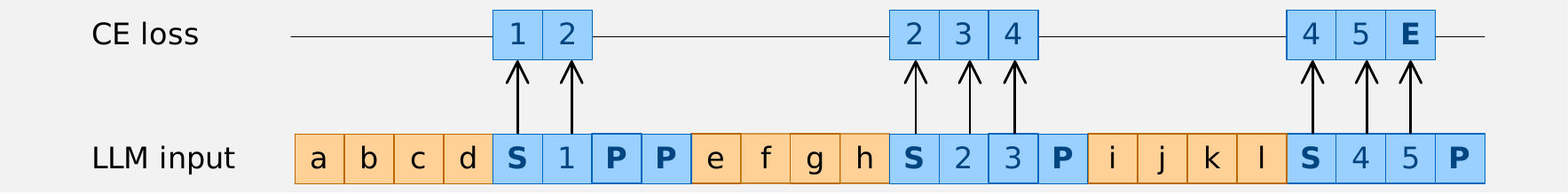}
\end{minipage}%
}%
\label{fig_train_paradigms}
\caption{Different training paradigms of Uni-ASR.}
\end{figure}

Following the training protocol of CosyVoice2~\cite{du2024cosyvoice}, speech and text tokens are interleaved according to a predefined \textit{N:M} ratio, and unoccupied text positions within each segment are filled with \verb+<pad>+ token. Figure~\ref{Fig2_1b} depicts the input sequence and its corresponding target transcription under the ~\textbf{SS} training paradigm. Note that while the figure illustrates a 1:1 ratio for clarity, our actual training configuration employs a 2:1 speech-text token ratio.

% (\textit{i.e.}, N frames of continuous speech representation followed by M text tokens).
% Note that our actual training configuration adopts a 2:1 speech-text token ratio, not the 1:1 ratio illustrated in the figure.

% To accommodate variations in speech rate and linguistic content, redundant text positions are padded with \verb+<pad>+ tokens.

% Adhering to the training protocol of CosyVoice2~\cite{du2024cosyvoice}, speech and text tokens are interleaved according to a predefined \textbf{N:M} ratio (i.e., N frames of continuous speech representation followed by M text tokens). To address mismatches arising from variations in speech rate, surplus text positions are padded with \verb++ tokens. Figure~\ref{Fig2_1b} depicts the input sequence and its corresponding target transcription under the standard streaming training paradigm. It should be noted that while the figure illustrates a 1:1 ratio for clarity, our actual training configuration employs a 2:1 speech-to-text token ratio.

% While efficient, this paradigm is prone to boundary transcription errors due to irreversible early token commitments.

% Online recognition systems that process input as fixed-length speech chunks can lead to unreliable transcriptions at the ends of blocks due to random truncation. 

For streaming inference, speech inputs arrive incrementally in chunks, the prefill–decode cycle repeats iteratively until the complete utterance is received. To formalize this workflow, we conceptualize it as a multi-turn dialogue paradigm implemented within non-streaming inference framework, where each speech chunk corresponds to a distinct dialogue turn. Crucially, key-value~(KV) caches computed during both prefill and decode stages are incrementally accumulated across chunks, while caches derived from prior chunks remain strictly immutable and are efficiently reused in subsequent processing. This mechanism ensures complete contextual continuity while eliminating redundant computations throughout the inference process.

% Critically, key-value~(KV) caches computed during both the prefill and decode stages are incrementally accumulated across chunks, the KV states from prior chunks remain immutable and are efficiently reused during subsequent chunk processing.

% KV states from previously processed chunks remain immutable and are consistently reused for subsequent chunks, thereby preserving full contextual continuity and avoiding redundant recomputation throughout the inference process.

\subsection{Context-aware Training Paradigm and Latest-Token Fallback Decoding Strategy}
\label{Context-aware Training and Latest-Token Fallback Decoding}
% Streaming automatic speech recognition (ASR) systems that process speech in fixed-length chunks often exhibit degraded transcription accuracy near chunk boundaries due to contextual discontinuities introduced by rigid segmentation. To mitigate this issue, Yang et al. proposed the Future-aware Transformer (FaT), which employs lookahead windows to propagate contextual cues from subsequent chunks into the current processing window, thereby strengthening contextual modeling. Complementing this approach, Zeghidour et al. introduced the Delayed Streaming Model (DSM), leveraging the wait-k strategy to flexibly regulate the accuracy–latency trade-off. While both methods significantly improve streaming recognition performance, they inherently introduce non-negligible latency overhead.

% In streaming ASR systems, recognition outputs are particularly prone to boundary-induced ambiguities and acoustic confusability due to insufficient contextual information—especially near the edges of fixed-length processing chunks. Although methods such as the Future-aware Transformer (FaT)\cite{yang2024learning} and Delayed Streams Modeling (DSM)\cite{zeghidour2025streaming} aim to mitigate these issues by enriching contextual modeling through look-ahead mechanisms, they inherently incur non-trivial latency penalties.

In streaming ASR systems, recognition outputs are particularly prone to boundary-induced ambiguities and acoustic confusability due to insufficient contextual information. Although Future-aware Transformer~(FaT)~\cite{yang2024learning} and Delayed Streams Modeling~(DSM)~\cite{zeghidour2025streaming} aim to mitigate these issues by enriching contextual modeling through look-ahead mechanisms, they inherently incur non-trivial latency penalties.

%Building upon the hold-n strategy, we propose a latest-token fallback decoding strategy: the most recently generated token of the current speech chunk is designated as a draft token and emitted in real time to ensure low-latency streaming inference. Upon receipt of the subsequent chunk, this draft token is invalidated and recomputed with updated contextual information to refine the final output. In the latest-token fallback decoding strategy, the initial token generated at each decoding step (beyond the first) is inherited from the preceding speech chunk rather than derived from the current input. This creates a train-inference mismatch: during inference, the model relies on cross-chunk contextual dependencies that are absent in conventional training setups. To align training dynamics with actual inference behavior, we introduce a context-aware training paradigm that explicitly models and leverages such inter-chunk contextual propagation. 

Building on the hold-\textit{n} strategy, we propose a latest-token fallback decoding strategy: the last text token of each speech chunk is emitted provisionally to ensure low-latency streaming and re-decoding upon processing the next chunk. However, this strategy introduces a train-inference mismatch—the first text token in subsequent decoding steps originates from the prior chunk, creating cross-chunk contextual dependencies unmodeled in conventional training. To bridge this gap, we introduce a context-aware streaming~(\textbf{CS}) training paradigm that can model cross-chunk contextual dependencies within continuously interleaved speech-text sequences.
% that explicitly learns and leverages inter-chunk contextual dependencies.

% As illustrated in Figure~\ref{Fig2_1c}, we designed a tailored variant of input and target sequences to align training with inference in streaming mode. Specifically:

% This deliberate modification not only prevents the training setup from degenerating into the standard streaming paradigm, but also ensures that the model can learn to recognize contextual incompleteness during training by intentionally masking the last text token in each input segment.

As depicted in Figure~\ref{Fig2_1c}, the \textbf{CS} training paradigm aligns training dynamics with inference behavior through two targeted sequence modifications: (i) input sequence: the last text token of each interleaved speech-text segment is replaced by \verb+<pad>+; (ii) target sequence: the \verb+<eos>+ is omitted from each text segment. This strategic formulation serves dual purposes. First, it prevents conflation with \textbf{SS} training paradigms. Second, by deliberately masking last text token during the training, the model learns to recognize cross-chunk contextual discontinuities and supplement missing text token. Consequently, during inference, when incremental decoding encounters a missing text token, the model triggers re-decoding from the designated context boundary. This mechanism explicitly bridges the train-inference gap by embedding context-aware re-decoding behavior directly into the learning objective.

% Second, by deliberately masking terminal tokens during training, the model learns to recognize contextual discontinuities and anticipate incomplete segment boundaries.  

%This deliberate modification not only prevents the training setup from collapsing into the standard streaming paradigm, but also ensures the model can learn to recognize contextual incompleteness during training by intentionally masking the last text token in each input segment.

%This deliberate modification not only prevents the training framework from degenerating into standard streaming paradigm, but also enables the model to learn to recognize contextual incompleteness during training through strategic masking of the final text token in each input segment.

%This deliberate modification prevents the training setup from collapsing into standard streaming paradigm. Crucially, by intentionally masking the latest text token in every input segment, the model learns during training to recognize contextual incompleteness. 
% generated by the previous round of decoding

For streaming inference based on the latest-token fallback decoding strategy, upon receiving a new speech chunk, the LLM discards the KV cache generated by the previous decoding round and resets the output sequence by replacing its last text token by \verb+<pad>+. The revised sequence is concatenated with the current speech representation as input to the prefill stage, enabling context-aware regeneration of the missing token in the current decoding round. Operating strictly within the current inference window, the strategy introduces negligible computational overhead while preserving stringent latency constraints.

\section{Experimental Setup}

\subsection{Data processing} 

% We evaluate Uni-ASR against several SOTA open-source ASR models across multiple widely adopted public benchmarks, including  AISHELL~\cite{bu2017aishell,du2018aishell}, LibriSpeech~\cite{panayotov2015librispeech}, FLEURS~\cite{conneau2023fleurs} and WeNetSpeech~\cite{zhang2022wenetspeech}. 

Uni-ASR is trained strictly on a Chinese–English bilingual corpus—no other languages are included. This focused corpus integrates open-source ASR datasets (AISHELL~\cite{bu2017aishell,du2018aishell}, WeNetSpeech~\cite{zhang2022wenetspeech}, FLEURS~\cite{conneau2023fleurs}, LibriSpeech~\cite{panayotov2015librispeech} and GigaSpeech~\cite{chen2021gigaspeech,yang2025gigaspeech}) with in-house labeled speech data. The hybrid composition substantially enriches intra-lingual diversity (regional accents, speaking styles, domain variations) and acoustic variability, thereby strengthening model robustness across diverse real-world conditions.

All text labels are subjected to text normalization, including case standardization, punctuation removal, and numeral-to-word conversion, to ensure constraints of streaming training paradigm. Character-level alignments,  obtained via MMS\_FA~\cite{pratap2024scaling}, are aggregated to the subword-token level to support BPE-based tokenization. For Chinese text, character-level segmentation is enforced by masking multi-character BPE-tokens, ensuring that each Chinese character is consistently treated as an independent linguistic unit during both encoding and alignment.

% All text labels undergo text normalization, including case standardization, punctuation removal and numeral-to-word conversion, to ensure constraints of streaming training paradigm. Character-level alignments derived via MMS\_FA~\cite{pratap2024scaling} are aggregated to the subword-token level to support BPE-based tokenization. Character-level word segmentation is enforced in Chinese text by masking Chinese BPE-tokens containing multiple characters. Thus, each Chinese character is encoded and aligned as an independent linguistic unit throughout the entire training and inference process.

% the input constraints of the streaming training paradigm.
% consistent training targets
% Following ConsyVoice2, character-level tokenization is enforced for Chinese text by excluding multi-character BPE tokens, thereby guaranteeing that each Chinese character is encoded and aligned as an independent linguistic unit throughout training and inference.

\subsection{Training of Audio Encoder}
We initialize a joint CTC/AED ASR model using the pre-trained weights of FireRedASR-AED and train it within the WeNet~\cite{yao2021wenet,zhang2022wenet} framework. During training, the encoder simultaneously employs full attention and dynamic chunk attention, enabling a unified framework for both non-streaming and streaming speech processing. To ensure strict causality in streaming mode, we replace the conventional depthwise convolutions in the Conformer blocks with causal convolutions.

% Throughout all subsequent supervised fine-tuning stages, we concurrently apply full attention and dynamic chunk attention mechanisms to process audio features, thereby preserving the audio encoder’s streaming capability.

% During all subsequent supervised fine-tuning stages, audio features are processed using both full attention and dynamic chunk attention mechanisms concurrently. This dual-attention strategy explicitly maintains the audio encoder’s streaming capability throughout training while preserving compatibility with non-streaming inference modes.

% In all subsequent supervised fine-tuning stages, audio features are processed simultaneously using both full attention and dynamic chunk attention mechanisms. This dual-attention strategy ensures the audio encoder retains its streaming capability throughout training while maintaining compatibility with non-streaming inference modes.

\subsection{Supervised Fine-tuning}

% The training of Index-ASR consists of three stages: audio encoder training, supervised fine-tuning. The training configuration of supervised fine-tuning are shown in Table.

% \subsubsection{Training of Audio Encoder}
% Following the approaches of FireRedASR and Index-ASR, we first train an ASR model based on a joint CTC/AED (Connectionist Temporal Classification / Attention-based Encoder-Decoder) architecture, which integrates a Conformer encoder with a Transformer decoder. During encoder training, both full-attention and dynamic chunk-attention are employed, enabling the model to operate in a unified framework that supports both non-streaming and streaming inference. Notably, to ensure strict causality in streaming mode, we replace the conventional depthwise convolutional modules in the Conformer block with causal convolutions, thereby maintaining temporal consistency in real-time scenarios.

%强调chunk-attention 和 full-attention 并行训练 U2架构的conformer-transformer 的CTC/AED模型，使得encoder具备处理流式音频的能力

As detailed in Table~\ref{SFT}, the supervised fine-tuning~(SFT) consists of five sequential stages, which are divided into non-streaming SFT and joint SFT according to their underlying training paradigm. Throughout all stages of SFT, the audio encoder simultaneously employs full attention and dynamic chunk attention mechanisms. This dual-attention strategy ensures that the encoder module maintains compatibility of speech processing modes throughout the training process.

% This dual attention strategy ensures that the encoder module maintains compatibility with speech processing modes throughout the training process.

%Throughout all SFT stages, both full attention and dynamic chunk attention mechanisms are employed simultaneously by the audio encoder to process speech, which ensures that the audio encoder consistently maintains mode‘s compatibility across the entire training process.
%Throughout all SFT stages, both full attention and dynamic chunk attention mechanisms are employed simultaneously by the audio encoder to process speech. This dual-attention strategy ensures that the audio encoder consistently maintains compatibility with both non-streaming and streaming speech processing modes across the entire training process.

% speech features are concurrently processed using both full attention and dynamic chunk attention mechanisms.

% Supervised fine-tuning (SFT) comprises five sequential stages, the training configuration of supervised fine-tuning are shown in Table 1.

\textbf{Non-streaming SFT}: It comprises four stages, with the training methodology for each stage adhering to well-established practices from mainstream LLM-based ASR frameworks. For further details, please refer to the Fun-ASR technical report~\cite{an2025fun}.

\textbf{Joint SFT}: Building upon the non-streaming SFT stage, stage-5 implements a unified joint training framework that integrates three distinct training paradigms: non-streaming, standard streaming, and context-aware streaming. These modes are uniformly sampled in a 1:1:1 ratio throughout training. Critically, standard and context-aware streaming paradigms are activated exclusively during training steps that employ the dynamic chunk attention mechanism, ensuring precise alignment between architectural capability and training dynamics.

% Building upon the non-streaming SFT phase, stage-5 incorporates interleaved speech-text input sequences with associated loss constraints, as detailed in Section~\ref{training paradigms} and Section~\ref{Context-aware Training and Latest-Token Fallback Decoding}. Training data is sampled uniformly (1:1:1) across three paradigms: non-streaming, standard streaming, and context-aware streaming. Notably, the streaming training paradigms are activated exclusively during training steps where dynamic chunk attention is utilized.

% Building on the non-streaming SFT phase, Stage 5 integrates interleaved text-speech input sequences with associated loss constraints, as detailed in Section~\ref{training paradigms} and Section~\ref{Context-aware Training and Latest-Token Fallback Decoding}. Training data is sampled uniformly (1:1:1) across three paradigms: non-streaming, standard streaming, and context-aware streaming. Notably, the streaming paradigms are activated exclusively during training steps where dynamic chunk attention is utilized.

% \includegraphics[height=1em]{fire.jpg}
% \includegraphics[height=1em]{snow.jpg} 

\begin{table} %[hptb]
\centering
\caption{Module configurations in multi-stage training. {\color{cyan}\faSnowflake} denotes module freezing, {\color{orange!90!red}\faFire} denotes full-parameter fine-tuning.}
\begin{tabular}{ccccc}
\toprule
\begin{tabular}[c]{@{}c@{}}SFT \\ Stage\end{tabular} & \begin{tabular}[c]{@{}c@{}}Audio-\\ Encoder\end{tabular} & Adapter & LLM & \begin{tabular}[c]{@{}c@{}}Training-\\ Paradigm\end{tabular} \\ 
\midrule
1         &{\color{cyan}\faSnowflake}                                                 &{\color{orange!90!red}\faFire}        &{\color{cyan}\faSnowflake}    & \textbf{NS}                                             \\
2         & {\color{orange!90!red}\faFire}                                                       &{\color{orange!90!red}\faFire}       &{\color{cyan}\faSnowflake}    & \textbf{NS}                                               \\
3         & {\color{cyan}\faSnowflake}                                                        & {\color{cyan}\faSnowflake}        &{\color{orange!90!red}\faFire}   & \textbf{NS}                                               \\
4         &{\color{orange!90!red}\faFire}                                                        &{\color{orange!90!red}\faFire}       &{\color{orange!90!red}\faFire}   & \textbf{NS}                                               \\
5         &{\color{orange!90!red}\faFire}                                                        &{\color{orange!90!red}\faFire}       &{\color{orange!90!red}\faFire}   & \textbf{NS/SS/CS}                                       \\ 
\bottomrule
\end{tabular}
\label{SFT}
\end{table}

\section{Experimental Results}
% We evaluate Index-ASR alongside several open-source speech recognition models on both publicly available ASR benchmarks and in-house test sets. For the open-source benchmarks, we conduct evaluations on the test sets of AISHELL~\cite{bu2017aishell,du2018aishell}, LibriSpeech~\cite{panayotov2015librispeech}, FLEURS~\cite{conneau2023fleurs} and WeNetSpeech~\cite{zhang2022wenetspeech}. Considering that most of the aforementioned test sets are relatively clean and contain limited background noise, we further assess the robustness of Index-ASR and other open-source models on internal test sets with complex background noise.

% We evaluate Uni-ASR against several SOTA open-source ASR models across multiple widely adopted public benchmarks, including  AISHELL~\cite{bu2017aishell,du2018aishell}, LibriSpeech~\cite{panayotov2015librispeech}, FLEURS~\cite{conneau2023fleurs} and WeNetSpeech~\cite{zhang2022wenetspeech}. 

% We evaluate Uni-ASR against several SOTA open-source ASR models across multiple widely adopted public benchmarks.  %including AISHELL, LibriSpeech, FLEURS and WeNetSpeech. 

% For recognition accuracy, we report either word error rates (WERs) or character error rates (CERs).

\subsection{Recognition Results}

%\sout{The non-Streaming results presented in Table~\ref{t2} indicate that all evaluated models exhibit strong recognition performance on the open-source test sets. However, on the noisy GigaSpeech test set, our model attains the best performance among the compared open-source systems, demonstrating its superior robustness under challenging acoustic conditions.}

% Table~\ref{t2} summarizes the non-streaming recognition results on public benchmarks, demonstrating that Uni-ASR achieves performance on par with SOTA models in offline ASR tasks.

% Table~\ref{t3} summarizes streaming recognition performance on AISHELL and Librispeech test sets. Uni-ASR outperforms native streaming models—Speech-RealmLLM, SpeechLLM-XL and Mocha-ASR—validating the efficacy of our streaming training paradigm. Although Qwen3-ASR-1.7B achieves the lowest WERs, its approach relies on iterative non-streaming decoding passes, incurring substantial computational overhead and resource inefficiency compared to native streaming architecture.

We evaluate Uni-ASR against several SOTA open-source ASR models across multiple widely adopted public benchmarks, including AISHELL, LibriSpeech, FLEURS and WeNetSpeech.

Table~\ref{t2} summarizes Uni-ASR’s recognition performance across non-streaming and streaming inference modes on established public benchmarks. The model attains non-streaming accuracy competitive with SOTA approaches while maintaining robust streaming capability—demonstrating the efficacy of its unified architecture.

%As detailed in Table~\ref{t3}, under the standard streaming inference framework, Uni-ASR consistently outperforms Speech-RealmLLM, SpeechLLM-XL and Mocha-ASR.

% Table~\ref{t3} further validates that Uni-ASR, trained  under the proposed joint training paradigm, surpasses native streaming frameworks: Speech ReaLLM, SpeechLLM-XL and MoCha-ASR in streaming recognition performance—without incorporating any auxiliary streaming-specific decoding techniques.  

Table~\ref{t3} further demonstrates that the proposed Uni-ASR outperforms native streaming frameworks, including Speech ReaLLM, SpeechLLM-XL and MoCha-ASR, in terms of streaming recognition performance without any auxiliary streaming-specific decoding techniques. Although Qwen3-ASR-1.7B achieves the lowest WER in streaming performance, its approach relies on iterative and repetitive non-streaming decoding passes, incurring substantial computational overhead compared to native streaming architecture.

\begin{table*}[t]
\centering
\caption{Comparison of ASR performance on open-source datasets. Results are reported in character error
rate (CER) and word error rate (WER), covering Chinese and English benchmark datasets respectively. For Uni-ASR, beam search decoding with a width of 3 is employed; the symbol $\ddagger$ represents the streaming recognition result with a chunk length of 1000 ms.} 
\label{t2}
\begin{tabular}{c|cccccccc}
\toprule
Test set            & \begin{tabular}[c]{@{}c@{}}GLM-ASR\\ -nano\end{tabular} & \begin{tabular}[c]{@{}c@{}}Whisper-\\ large-v2\end{tabular} & \begin{tabular}[c]{@{}c@{}}Seed-\\ ASR\end{tabular}  & \begin{tabular}[c]{@{}c@{}}FireRedASR\\ -AED\end{tabular} & \begin{tabular}[c]{@{}c@{}}Fun-ASR\\ -nano\end{tabular} & \begin{tabular}[c]{@{}c@{}}Qwen3-ASR\\ -1.7B\end{tabular} & \begin{tabular}[c]{@{}c@{}}Uni-\\ ASR\end{tabular}  & \begin{tabular}[c]{@{}c@{}}Uni-\\ ASR$^\ddagger$\end{tabular}\\ \midrule
AISHELL-1                 & 1.81                                                    & -                                                           & 0.68                                                & 0.55                                                      & 1.80                                                    & -                                                                             & 1.44                                               & 2.15                                                          \\
AISHELL-2                 & -                                                       & -                                                           & 2.27                                                & 2.52                                                      & 2.75                                                    & 2.71                                                                          & 2.60                                               & 3.25                                                          \\
LibriSpeech test-clean    & 2.00                                                    & 2.5                                                         & 1.58                                                & 1.93                                                      & 1.76                                                    & 1.63                                                                          & 1.93                                               & 2.44                                                          \\
LibriSpeech test-other    & 4.19                                                    & 4.9                                                         & 2.84                                                & 4.44                                                      & 4.33                                                    & 3.38                                                                          & 4.11                                               & 5.71                                                          \\
FLEURS\_zh                & -                                                       & 14.7                                                        & -                                                   & -                                                         & 2.56                                                    & 2.41                                                                          & 2.57                                               & 3.43                                                          \\
FLEURS\_en                & -                                                       & 4.2                                                         & 3.43                                                & -                                                         & 5.96                                                    & 3.35                                                                          & 4.44                                               & 6.24                                                          \\
WeNetSpeech Test\_Meeting & 6.73                                                    & -                                                           & 5.69                                                & 4.76                                                      & 6.60                                                    & 5.88                                                                          & 6.32                                               & 8.04                                                          \\
WeNetSpeech Test\_Net     & -                                                       & -                                                           & 4.66                                                & 4.88                                                      & 6.01                                                    & 4.97                                                                          & 5.78                                               & 6.44                                                          \\
 
\bottomrule
\end{tabular}
\end{table*}

% \subsection{Streaming Results}

% Our proposed fallback decoding strategy can improve the accuracy of stream cytometry recognition by correcting unreliable boundary transcriptions. 

% Our proposed fallback decoding strategy can improve the accuracy of stream cytometry recognition by correcting unreliable boundary transcriptions. 

% Our proposed fallback decoding strategy can improve the accuracy of stream cytometry recognition by correcting unreliable boundary transcriptions. 

% Our proposed fallback decoding strategy can improve the accuracy of stream cytometry recognition by correcting unreliable boundary transcriptions. 

% Our proposed fallback decoding strategy can improve the accuracy of stream cytometry recognition by correcting unreliable boundary transcriptions. 

% Comparison of streaming ASR performance on test sets. Results are reported in CERs(\%) for AISHELL and WERs(\%) for Librispeech.

% $2.46 \mid 3.54$
\begin{table} %[!htbp]
\centering
\caption{Comparison of streaming ASR performance on test sets. Results are reported in CER for AISHELL and WER for Librispeech.} 
\label{t3}
{
\setlength{\tabcolsep}{5pt}
\begin{tabular}{c|c|cc}
\toprule
Model               & \begin{tabular}[c]{@{}c@{}}Chunk \\ length\end{tabular} & AISHELL   & LibriSpeech \\ \midrule
% Transducer-Llama    & 160           &  -      & $2.5 \mid 6.5$   \\
Speech ReaLLM      & 1920          &  -      & $3.0 \mid 7.4$     \\
SpeechLLM-XL        & 1280          &  -      & $2.7 \mid 6.7$     \\
%CTC+Context prompts & -             &  -      & $3.2 \mid 7.9$     \\
%BTI                 & -             & $5.9 \mid 7.2$   &  -        \\
MoCha-ASR           & -             & $5.1 \mid 5.5$   &  -      \\ 
Qwen3-ASR-1.7B      & 1000          & -   & $1.95 \mid 4.51$      \\ \midrule
Uni-ASR             & 1000          & $2.15 \mid 3.25$ & $2.44 \mid 5.71$   \\ \bottomrule
\end{tabular}
}
\end{table}

\subsection{Ablation Study}

We performed systematic ablation studies to quantitatively assess how chunk length and decoding strategy variations influence streaming recognition performance.

%on AISHELL-1, AISHELL-2, and LibriSpeech (test\_clean / test\_other)

\subsubsection{Chunk Length}
In chunk-based streaming speech recognition systems, reducing the length of speech chunk can reduce latency but incurs a significant performance degradation—this is a fundamental latency-accuracy trade-off inherent to streaming architectures.

As demonstrated in Table~\ref{table_ablation}, under beam search decoding, progressively reducing the length of speech chunk from the baseline of 1000 ms to 640 ms and 320 ms incurs relative average WER increases of 10.02\% and 32.15\%, respectively. The pronounced degradation in streaming performance is primarily attributable to the increased susceptibility of shorter data chunks to fragmentation artifacts and boundary transcription errors, which are caused by insufficient contextual information.

% This substantial degradation in streaming inference performance directly stems from the heightened vulnerability of shorter chunks to boundary-induced transcription errors and fragmentation artifacts at chunk transitions.

% As evidenced in Table~\ref{table_ablation}, using the standard beam search decoding method, reducing the speech chunk length from the 1000 ms baseline to 640 ms and 320 ms results in relative increases of 10.02\% and 32.15\% in the average WER, respectively. Clearly, shorter speech chunks induce a pronounced degradation in streaming inference performance, as they are more susceptible to boundary-induced transcription errors and fragmentation artifacts at chunk transitions.

% compared to larger chunks.

\subsubsection{Decoding strategy}

%To alleviate the aforementioned issues, we propose a latest-token fallback decoding strategy that improves streaming recognition performance through delay correction.

To address the aforementioned challenges, we propose a novel latest-token fallback decoding strategy that enhances streaming recognition performance by leveraging delay correction.

As showed in Table~\ref{table_ablation}, the fallback decoding strategy yields gains in recognition performance across a spectrum of speech chunk lengths, with the most substantial benefits observed at shorter chunk lengths. At chunk lengths of 1000 ms, 640 ms and 320 ms, this strategy achieves relative reductions in average WER of 11.43\%, 13.50\% and 19.88\% over standard greedy decoding, respectively, while consistently matching or exceeding the performance of beam search decoding across same evaluated conditions. Collectively, these results demonstrate that the fallback decoding strategy delivers enhanced streaming decoding accuracy and optimized inference efficiency under real-world deployment conditions, thereby underscoring its exceptional suitability for latency-sensitive and resource-constrained applications.

\begin{table}[t] %[!htbp]
\centering
\caption{Results are reported ablation study of streaming performance across diffent chunk lengths and decoding methods. Specifically, “Beam” denotes the beam search decoding method with a width of 3, and the symbol $\dagger$ denotes the proposed latest-token fallback decoding strategy is employed.} 
\label{t_ablation}
{
\setlength{\tabcolsep}{5pt}
\begin{tabular}{c|c|ccc}
\toprule
\begin{tabular}[c]{@{}c@{}}Chunk \\ length\end{tabular} & \begin{tabular}[c]{@{}c@{}}Decoding \\ method\end{tabular} & AISHELL  & LibriSpeech & \textit{\textbf{Avg}}  \\ \midrule
\multirow{4}{*}{1000}                                    & Greedy & $2.46 \mid 3.54$ & $2.74 \mid 6.65$                                                         & 3.85 \\
                                                         
                                                         & Greedy$^\dagger$     & $2.14 \mid 3.28$ & $2.40 \mid 5.82$      & 3.41 \\
                                                         
                                                         & Beam   & $2.15 \mid 3.25$         & $2.44 \mid 5.71$      & 3.39      \\
                                                         & Beam$^\dagger$   & $1.91 \mid 3.04$      & $2.22 \mid 5.30$  & 3.12 \\ 
                                                         \midrule
                                                         
\multirow{4}{*}{640}                                     & Greedy & $2.63 \mid 3.92$ & $2.99 \mid 7.32$                                                         & 4.22 \\
                                                         
                                                         & Greedy$^\dagger$     & $2.36 \mid 3.35$ & $2.52 \mid 6.37$      & 3.65 \\
                                                         
                                                         & Beam   & $2.35 \mid 3.61$         & $2.66 \mid 6.31$      & 3.73      \\
                                                         & Beam$^\dagger$   & $2.09 \mid 3.10$      & $2.32 \mid 5.68$  & 3.30 \\
                                                         \midrule
                                                         
\multirow{4}{*}{320}                                     & Greedy & $3.31 \mid 4.51$ & $3.67 \mid 9.22$                                                         & 5.18 \\
                                                         
                                                         & Greedy$^\dagger$     & $2.61 \mid 3.75$ & $2.75 \mid 7.50$      & 4.15 \\
                                                         
                                                         & Beam   & $2.90 \mid 4.11$         & $3.21 \mid 7.71$      & 4.48      \\
                                                         & Beam$^\dagger$   & $2.38 \mid 3.52$      & $2.55 \mid 6.61$  & 3.77 \\
                                                         \bottomrule
                                                         
\end{tabular}
}
\label{table_ablation}
\end{table}

% \begin{tabular}{ccc}
% \toprule
% A & B & C \\
% \midrule
% 1 & 2 & 3 \\
% 4 & 5 & 6 \\
% \bottomrule
% \end{tabular}

\section{Conclusions and Discussions }

In this paper, we propose Uni-ASR, which establishes a unified LLM-based ASR framework that natively supports non-streaming and streaming inference within a single architecture. It not only achieves competitive non-streaming accuracy, but also demonstrates SOTA streaming performance across diverse latency budgets. Its joint training paradigm, context-aware streaming paradigm and fallback decoding strategy ensure strict training–inference consistency, ablation studies validate the critical roles and effectiveness of these components. This work provides a practical, resource-efficient solution for versatile ASR deployment and advances unified methodology bridging non-streaming accuracy and streaming efficiency.

\clearpage

\section{Generative AI Use Disclosure}
% The extent of Generative AI use must be disclosed. This section may be in the 5th or 6th pages of regular papers, or the 9th or 10th pages of long papers.  ISCA policy says: \textit{All (co-)authors must be responsible and accountable for the work and content of the paper, and they must consent to its submission. Any generative AI tools cannot be a co-author of the paper. They can be used for editing and polishing manuscripts, but should not be used for producing a significant part of the manuscript}.

Generative AI tools were employed only for linguistic refinement and grammatical polishing of the manuscript. All core scientific content—including research conception, methodology, experimental results, analysis, and conclusions—was independently developed, validated, and authored by the human research team, who retain full responsibility for integrity and accuracy of manuscript.

\bibliographystyle{IEEEtran}
\bibliography{mybib}

\end{document}